\documentclass[11pt, letterpaper]{article}

\usepackage[letterpaper,
            top=1.5in,
			bottom=1in,
			left=1in,
			right=1in,
			marginparwidth=1in]{geometry}
			
\usepackage[round]{natbib}
\usepackage{hyperref} 

\usepackage[english]{babel}
\usepackage[utf8]{inputenc}
\usepackage[T1]{fontenc}

\usepackage{geometry}
\usepackage{amsmath}
\usepackage{amssymb}

\usepackage{url}

\usepackage{fancyhdr} 

\usepackage{graphicx}
\usepackage{caption}
\usepackage{subcaption}

\usepackage{longtable}

\defcitealias{medpac}{MedPAC, 2007} 
\defcitealias{valueprograms}{CMS, 2017}
\defcitealias{hrrp}{CMS, 2011}

\title{Categorical Co-Frequency Analysis: Clustering Diagnosis Codes to Predict Hospital Readmissions}

\author{Hallee E. Wong
        \thanks{hew1@williams.edu} \thanks{Williams College, Williamstown, MA, USA}
        \and Brianna C. Heggeseth
        \thanks{Macalester College, Saint Paul, MN, USA} 
        \and
        Steven J. Miller\footnotemark[2]
        }

\date{}

\begin{document}

\maketitle

\begin{abstract}
Accurately predicting patients' risk of 30-day hospital readmission would enable hospitals to efficiently allocate resource-intensive interventions. We develop a new method, Categorical Co-Frequency Analysis (CoFA), for clustering diagnosis codes from the International Classification of Diseases (ICD) according to the similarity in relationships between covariates and readmission risk. CoFA measures the similarity between diagnoses by the frequency with which two diagnoses are split in the same direction versus split apart in random forests to predict readmission risk. Applying CoFA to de-identified data from Berkshire Medical Center, we identified three groups of diagnoses that vary in readmission risk. To evaluate CoFA, we compared readmission risk models using ICD majors and CoFA groups to a baseline model without diagnosis variables. We found substituting ICD majors for the CoFA-identified clusters simplified the model without compromising the accuracy of predictions. Fitting separate models for each ICD major and CoFA group did not improve predictions, suggesting that readmission risk may be more homogeneous that heterogeneous across diagnosis groups. 
\end{abstract}

\section{Introduction}

Frequent hospitalizations have the potential to negatively affect patients' health and strain healthcare systems' resources. Among Medicare patients, 17.6\% of hospitalizations result in readmission to in-patient care within 30 days of discharge (``30-day readmission'') at an annual cost of \$15 billion \citepalias{medpac}. Through the Affordable Care Act in 2012, the Centers for Medicare and Medicaid Services (CMS) incentivises American hospitals to reduce readmission rates by deducting up to 3\% of reimbursement from hospitals with higher than expected readmission rates \citepalias{hrrp}.
 
While many readmissions are unavoidable, some are preventable through discharge planning, follow-up case management, and patient education \citep{benbassat}. In randomized control trials, high intensity interventions such as a home visit by a registered nurse within 3 days of discharge, coordination with primary care providers, and individual case management have been shown to reduce readmission rates \citep{vergaegh}. Because the most effective interventions are resource intensive, statistical models for predicting patients' readmission risk are highly valuable for optimizing the allocation of hospital resources \citep{stukel}. 

\paragraph{Clinical Relevance}

Existing readmission risk models fail to utilize high dimensional diagnosis data available in electronic medical records (EMRs) in a reproducible or strategic way. At the most specific level, the 10$^{\text{th}}$ edition of the International Classification of Disease (ICD), has 68,000 unique codes, making the creation of predictive diagnosis features a non-trivial task. Previous studies either used diagnoses hand-picked by clinicians or considered every medical diagnosis satisfying a minimum frequency in the dataset \citep{lee, halfon, choletti, futoma, yu}. The ICD system is a hierarchy for coding diagnoses that is optimized for semantic clarity and efficient billing. Our first goal is to create a hierarchical clustering that groups diagnoses with similar readmission risk after accounting for other covariates. Such a hierarchy would allow clinicians to identify patients at risk of readmission by diagnosis and could identify subgroups of diagnoses which have significantly different outcomes from the general population and may need a particular risk model or require development of further targeted interventions.

\paragraph{Technical Significance}

Random forests have been used for feature selection by measuring variable importance as the average improvement in node purity. In this paper we consider how the categorical diagnosis variable splits to improve node purity of readmission risk. If two diagnoses have similar readmission risk, then we expect those diagnoses to frequently split together when the categorical diagnosis variable is used as the split variable in random forests. To identify diagnoses which are similar, and thus can be combined into one category to reduce the distinct levels of the primary diagnosis predictor, we analyze how frequently diagnoses are grouped together when the categorical diagnosis variable is used as the split criteria in random forests to predict readmission risk. In a new procedure called Categorical Co-Frequency Analysis (CoFA), we calculate a co-frequency statistic for each pair of diagnoses in a random forest and use those statistics to a form a distance matrix for hierarchical clustering. In this paper, we apply CoFA to data from Berkshire Medical Center to cluster diagnosis codes and then evaluate the clustering by using the identified groups as features in predictive models. 

Our second goal is to compare the performance of clusterwise logistic regression models to assess whether fitting separate models for diagnosis-defined subgroups results in more accurate predictions. If there exist subgroups within the population of hospitalizations with different relationships between predictors and the outcome then fitting clusterwise models stratified on those subgroups should perform better than a single global model. We assess the clusters identified by CoFA by comparing the predictive performance of those diagnosis features in both simple logistic regression models and clusterwise logistic regression models. 

\section{Cohort}
Berkshire Medical Center (BMC) is a medium-sized non-profit teaching hospital in rural western Massachusetts. Electronic medical records describing 19,720 in-patient hospital visits at BMC from September 1$^\text{st}$, 2015 to December 31$^\text{st}$, 2016 were extracted and de-identified in accordance with the Health Insurance Portability and Accountability Act (HIPAA) Privacy Rule. The data contained variables describing the patient's demographic information, hospital utilization and clinical diagnosis in the form of ICD codes. 

\subsection{Cohort Selection} 
All hospitalizations ending after December 1$^\text{st}$, 2016 were excluded from analysis because the patient's 30-day readmission status was right censored. For patients hospitalized multiple times during the study period, we considered each hospitalization as an independent observation. To include a predictor variable quantifying the number of hospitalizations in the previous 30 days without missing data, we excluded all hospitalizations starting before October 1$^\text{st}$ 2015. One patient had two overlapping hospital stays; we assume this was due to a clerical error and excluded the 11 hospitalizations involving that patient. Visits without any clinical diagnoses were also excluded, resulting in a final sample of $17,093$ hospitalizations and $10,895$ unique patients. 

\subsection{Outcome}
A patient was considered readmitted if after being discharged from an (index) inpatient stay at BMC they were readmitted to inpatient care \emph{again} at BMC within 30 days of the initial discharge date. Time between hospitalizations -- measured from the discharge date of the index visit to the admission date of the follow-up visit -- was used to create a binary indicator for 30-day readmission. 

\subsection{Non-diagnosis Features} 
The data included demographic characteristics of the patient (age, sex), logistical details of their admittance (admitted through emergency department, admission source) and discharge (length of stay, discharge disposition), the services they utilized during their stay (received surgery, number of medications, number of auxiliary diagnoses) and previous hospital utilization (number of hospitalizations in the previous 30 days). All categorical variables were converted to binary indicator variables. 

\subsection{Diagnosis Features}
Each row, describing a patient's hospital stay at BMC, included an ICD code for the patient's primary diagnosis. The ICD hierarchy consists of leaves (individual codes), majors, sub-chapters and chapters, in order of increasing generality (examples can be found in Appendix Figure \ref{fig:icd10}). All primary diagnosis codes were generalized to the ``major'' level to condense the categories while still preserving recognizable specific medical conditions. Lastly, all major diagnosis categories with fewer than 100 appearances were grouped into an `Other' category, resulting in 34 observed ``major'' diagnoses.

\section{Methods}

\subsection{Clinical Feature Creation} 

To identify groups of diagnoses with similar (potentially non-linear) relationships between predictors and the outcome of interest, we developed a method called Categorical Co-Frequency Analysis (CoFA). CoFA that measures the similarity between levels (i.e. categories) of a high dimensional categorical variable by the frequency with which the levels split in the same direction versus split apart in a random forest and uses this similarity measurement to create a distance matrix for use in clustering. 

\begin{figure}[htpb]
	\centering
	\begin{subfigure}[b]{0.5\textwidth}
        \includegraphics[width=\textwidth]{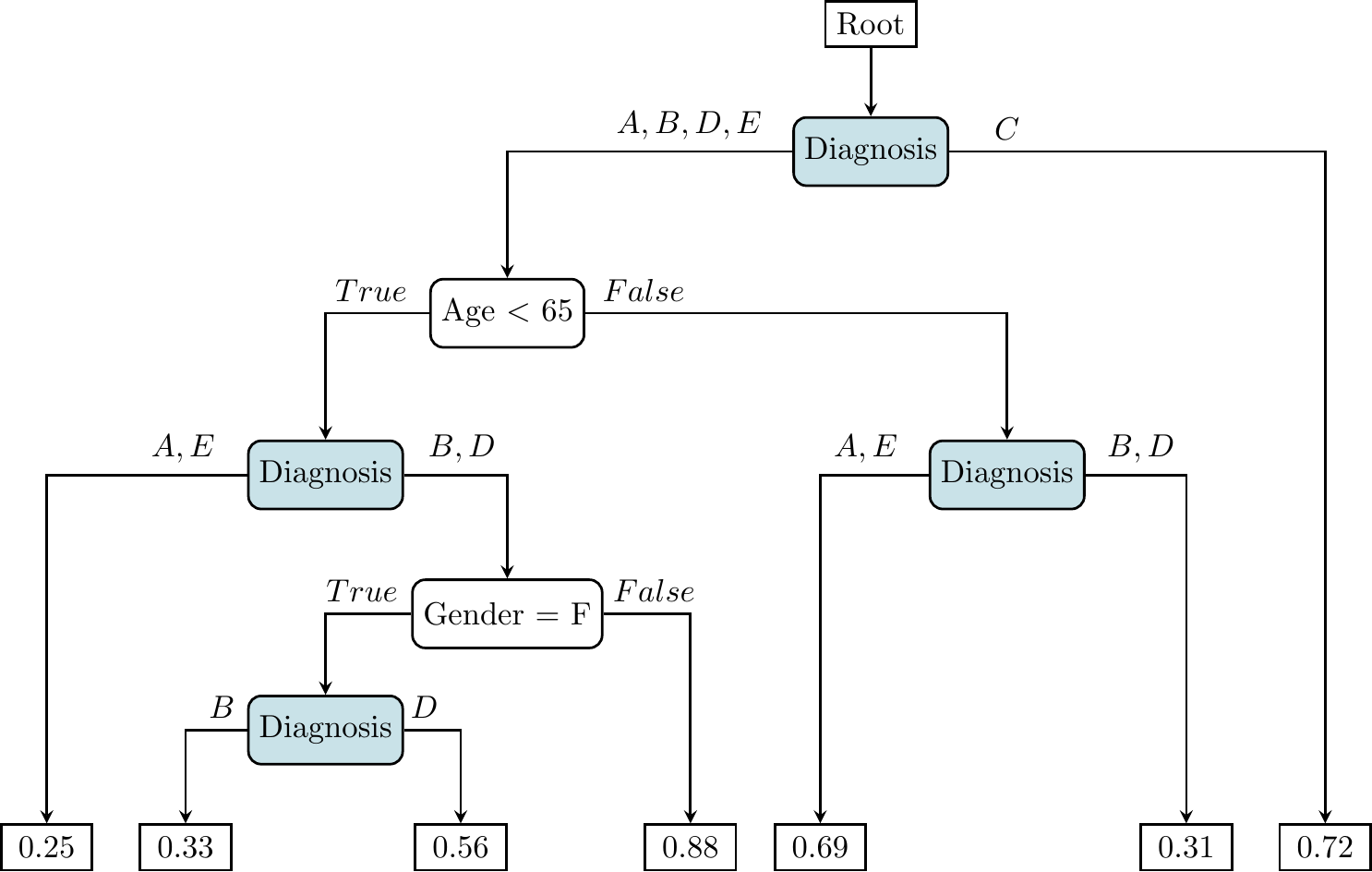}
	    \caption{An example decision tree fit on simulated data. Nodes splitting on diagnosis are highlighted.}
	\label{fig:ex_tree}
	\end{subfigure}
	\quad  ~
	\begin{subfigure}[b]{0.35\textwidth}
	    \includegraphics[width=\textwidth]{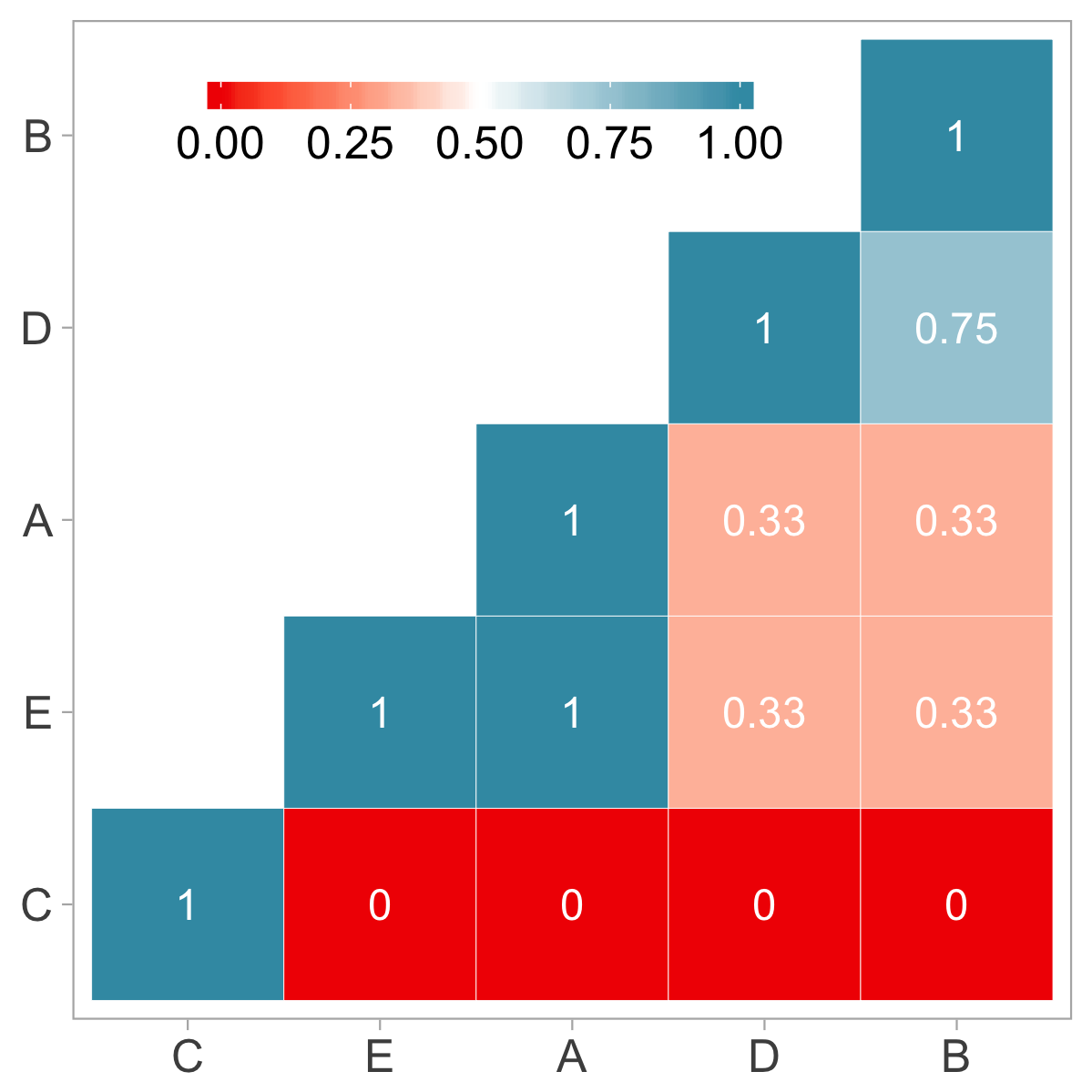}
	    \caption{Each cell shows the CoFA statistic for a pair of diagnoses.}
	    \label{fig:ex_matrix}
	\end{subfigure}
	\caption{An example decision tree and the corresponding matrix of CoFA statistics. The CoFA statistics quantify the frequency with which two levels of a categorical variable split together vs. split apart.}
	\label{fig:example}
\end{figure}

To perform CoFA, we first fit a random forest of Classification and Regression Trees (CART)\footnote{We implemented the random forests for CoFA using the \texttt{rpart} package for classification trees in R \citep{rpart, rlanguage}. When fitting the random forest, ''dice rolling'' was used to select the predictors considering for splitting at each node; for each of the $k$ variables available at a node a die with numbers $1$ through $k$ is rolled and if the value is $\leq \sqrt{k}$ then the variable is considered for splitting.} to predict the outcome of interest (e.g. 30-day readmission indicator) using all available predictors including the 34-category predictor (e.g. ICD major of primary diagnosis) we wish to cluster \citep{cart_book}. Every time the categorical predictor is used to split at a node in a decision tree within the forest, the levels of the categorical variable are partitioned into two groups to define the split. For level $i$ and level $j$ of the categorical predictor, we define the categorical co-frequency statistic or CoFA statistic as
\begin{align}
	s_{i,j} = \frac{\text{No. of times level $i$ and level $j$ split in the same direction}}{\text{No. times level $i$ and level $j$ are used to split at a node}} 
\end{align}
where the numerator and denominator are the totals across all trees in the random forest. If $s_{i,j}=1$, that indicates a strong similarity; if $s_{i,j}=0$, that indicates strong dissimilarity. If $s_{i,j}=0.5$, no clear relationship is indicated. We calculate the CoFA statistic for every pair of levels in the categorical predictor. Figure \ref{fig:example} illustrates an example of a matrix of co-frequency statistics calculated from a single decision tree.

Next, we eliminate co-frequency statistics that are not statistically significant. If two labels are interchangeable then there should be no systematic pattern in how they split at nodes, and $s_{i,j}$ should be close to $0.5$. We wish to test the null hypothesis ($H_0$) that level $i$ and level $j$ are interchangeable labels. To empirically determine the distribution of $s_{i,j}$ when $H_0$ is true we train 500 random forests fit on data with the levels of the categorical variable randomly shuffled and calculate $s_{i,j} \ \forall i,j$. Because the diagnoses are not evenly distributed, the null hypothesis distribution $s_{i,j}$ is not the same for every combination of diagnoses. For each pair of levels $i$, $j$ we calculate a z-score using the null hypothesis distribution and a p-value to test whether the observed $s_{i,j}$ was statistically significantly different from the distribution of values for $s_{i,j}$ when $H_0$ is true. For each CoFA statistic the z-score is defined  as
\begin{align}
    z = \frac{s_{i,j} - \mu_{i,j}}{\sigma_{i,j}} \ ,
\end{align}
where $\mu_{i,j}$ and $\sigma_{i,j}$ are the mean and standard error of the empirical null hypothesis distribution of $s_{i,j}$. We observed that under $H_0$ the statistic was approximately normally distributed, so to calculate p-value we compared $z$ to a standard normal distribution. To address the multiple testing problem -- testing the $s_{i,j}$ statistic for every pairwise combination of the 34 levels involves 561 tests -- we used a Bonferroni correction. The cutoff $\alpha=0.05/561$ translates to rejecting the null hypothesis if the magnitude of the z-score of $s_{i,j}$ is larger than $3.92$. 

To create a distance matrix for hierarchical clustering we re-scaled each $s_{i,j}$ to $d_{i,j} = 1-s_{i,j}$ so that $0$ corresponds no distance between groups and $1$ corresponds maximum distance between groups. For CoFA statistics that were not statistically significant, we set $s_{i,j} = 0.5$ such that $d_{i,j}=0.5$, a value at the center of $[0,1]$ the range of possible distances. Complete-linkage agglomerative hierarchical clustering was performed using the distance matrix of $d_{i,j}$ values to create a dendrogram relating categories by their co-frequency. We cut the tree to get a hard clustering of levels. 

\subsection{Model Fitting}

To predict probability of 30-day readmission we used logistic regression with LASSO feature selection \citep{glmnet}. Three models were fit considering different sets of predictors: 
\begin{enumerate}
    \item a baseline model considering all non-diagnosis variables
    \item a model considering all non-diagnosis variables and binary indicators for ICD majors
    \item a model considering all non-diagnosis variables and binary indicators for CoFA groups.
\end{enumerate}
For each model, the tuning parameter $\lambda$ was chosen following the 1 standard error rule using 5-fold cross validation on the training data. After choosing the tuning parameter and fitting a lasso model, we used the list of predictors with non-zero coefficients to refit a logistic regression with maximum likelihood estimation. To perform binary classification a probability cutoff can be chosen that satisfies the user's constraints on sensitivity or specificity. 

We also fit clusterwise logistic regression models in which a separate logistic regression model was fit with LASSO feature selection for each subgroup defined by a categorical stratification variable. Two clusterwise models were fit and compared to the baseline logistic regression model:
\begin{enumerate}
    \item a clusterwise model stratified by ICD major
    \item a clusterwise model stratified by CoFA group.
\end{enumerate}
To classify patients, separate probability cutoffs can be chosen for each submodel to satisfy constraints on sensitivity or specificity by subgroup. If there exist subgroups with significantly different relationships between predictors and the outcome then fitting a clusterwise model stratified by those subgroups should perform better than a single global model.

\subsection{Evaluating Model Success}
For a readmission model to be useful in a clinical setting, it should make rank accurate predictions, meaning that patients who are observed to be readmitted should have a higher predicted probability of readmission than those who are not. Rank accuracy was assessed using area under the receiver operator curve (AUC). AUC directly quantifies rank accuracy because it can be interpreted as the probability that a true positive example is assigned a higher predicted probability than a true negative example \citep{tilaki}. 

To compare clusterwise models to the baseline model, we used a weighted AUC. To calculate the weighted AUC for a model, we first calculate the AUC separately using the predicted and observed values of each subgroup (e.g., ICD majors, CoFA groups) and then take an average of the AUCs weighted by the number of observations in each subgroup. Calculating an overall AUC by pooling the predictions and observations from all subgroup model would imply that the same probability cutoff is used for all subgroup models when making predictions. By using a weighted AUC to assess models, we do not assume that all subgroup models must use the same probability cutoff to classify new data; different probability cutoffs may be chosen for the submodels so that each subgroup has the same sensitivity or specificity.

Repeated random sub-sampling was used to estimate predictive accuracy. For each model fitting procedure, 100 iterations were performed fitting a model on a randomly selected 80\% of observations and testing on the held out 20\% to estimate the out-of-sample AUC or weighted AUC. To determine whether the out-of-sample AUCs for one procedure were significantly different from those produced by another procedure across the iterations of repeated random sub-sampling, we use the corrected resampled (paired) t-test \citep{nadeau}. Let $x_i=a_i-b_i$ be the difference in test AUC for procedure A and procedure B in iteration $i$, then the test statistic is
\begin{align}
	t = \frac{\frac{1}{m} \sum_{i=1}^m x_i}{\hat{\sigma} \sqrt{\frac{1}{m} + \frac{n_{text}}{n_{train}}}} 
\end{align}
where $m$ is the number of iterations of repeated random sub-sampling and $n_{test}$ and $n_{train}$ are the number of observations in the test sets and training sets respectively. This test statistic is similar to that of a paired t-test except $\hat{\sigma}^2/m$ has been replaced with $\hat{\sigma}^2(\frac{1}{m}+\frac{n_{test}}{n_{train}})$ in the denominator to correct for the random overlap among training data and among testing data across the iterations. To test the set of hypotheses $H_0: \bar{x} = 0$, $H_A: \bar{x} \neq 0$ we compare $t$ to a student t-distribution with $m-1$ degrees of freedom. 

\section{Results} 

\subsection{CoFA Diagnosis Clusters}

\begin{figure}[bht]
    \centering 
    \begin{subfigure}[b]{0.45\textwidth}
        \includegraphics[width=\linewidth]{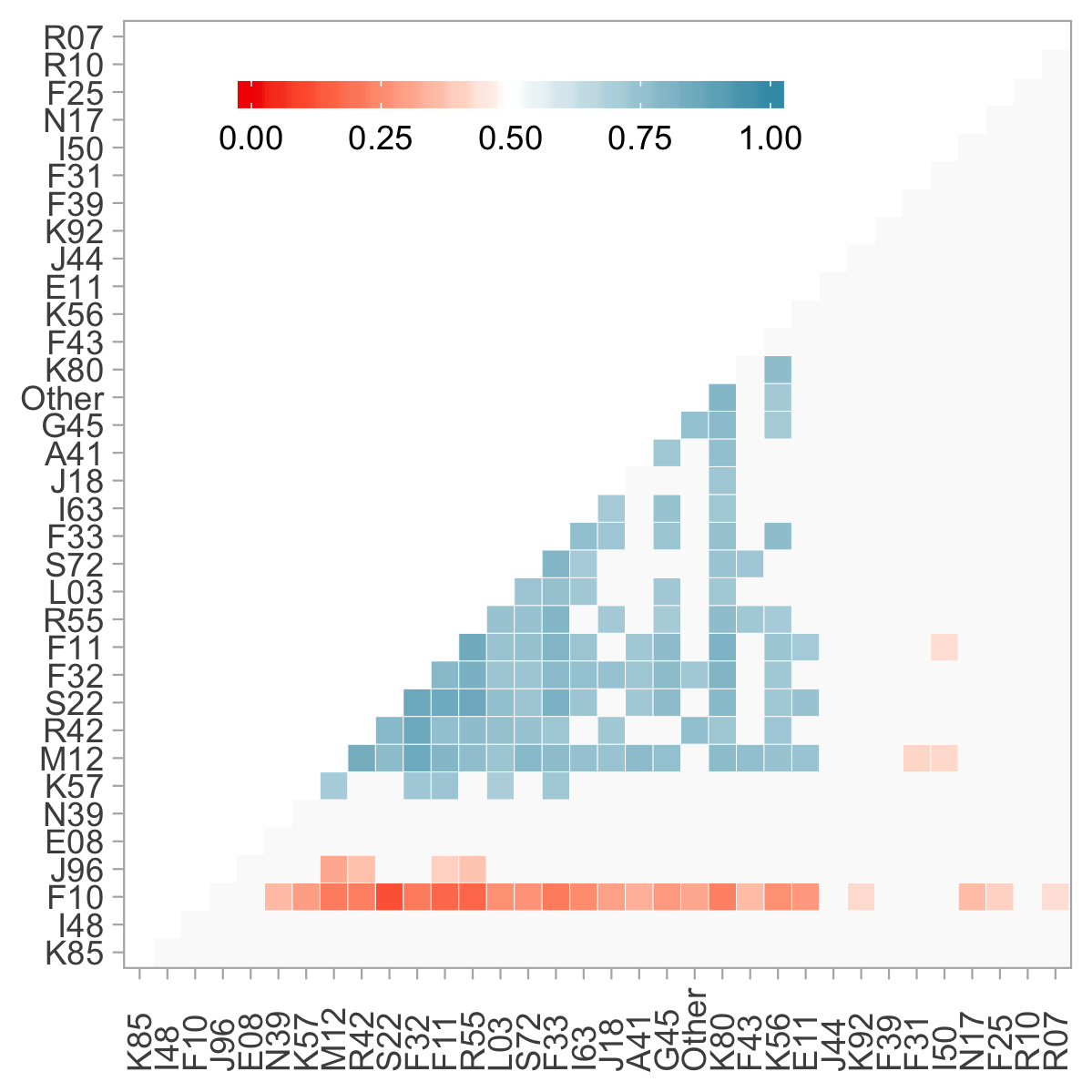}
        \caption{The CoFA statistic was calculated for each pair of ICD majors. Pairs of levels for which the CoFA statistic did not satisfy $|z|>3.92$ are shown in grey.}
        \label{fig:ccfa_matrix_masked}
    \end{subfigure}
    ~
    \begin{subfigure}[b]{0.45\textwidth}
        \includegraphics[width=\linewidth]{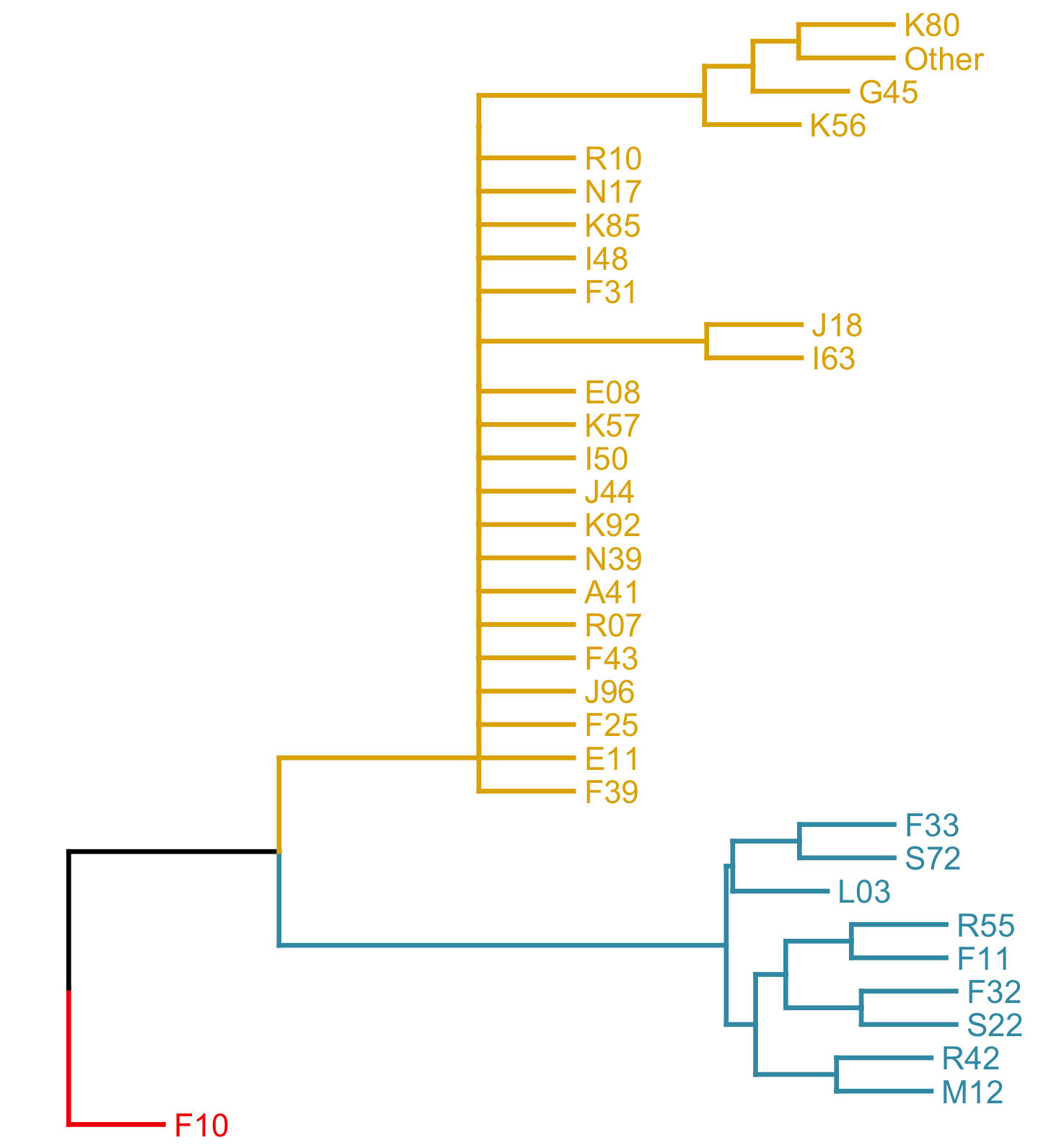}
        \caption{Hierachical clustering was performed using the values from Figure \ref{fig:ccfa_matrix_masked} as a distance matrix.}
        \label{fig:ccfa_hierarchy}
    \end{subfigure}
    \caption{CoFA on a random forest of 100 trees identified three main groups of diagnoses:  group 1 (yellow), group 2 (red) and group 3 (blue). Appendix Table \ref{tbl:ccfa_groups} provides descriptions of the ICD majors.}
\end{figure}

Figure \ref{fig:ccfa_matrix_masked} shows the CoFA statistics calculated for each pair of ICD majors in a random forest of 100 trees using the entire dataset. Pairs of diagnoses for which the CoFA statistic was not statistically significant ($|z| \leq 3.92$) are gray in \ref{fig:ccfa_matrix_masked} and were set to $0.5$ prior to performing hierarchical clustering (Figure \ref{fig:ccfa_hierarchy}). 

For greater interpretability, we cut the hierarchical clustering dendrogram (Figure \ref{fig:ccfa_hierarchy}) to create three groups, which are summarized in Table \ref{tbl:ccfa_stats}. The three CoFA groups loosely correspond to the pairs of ICD majors that neither have high or low co-frequency with others (Group 1), the one ICD major that has low co-frequency with most other majors (Group 2), shown in red, and the pairs of ICD majors in blue (Group 3), indicating high co-frequency, in Figure \ref{fig:ccfa_matrix_masked}. Group 2, containing only alcohol related diagnoses (F10), had the highest risk with an observed readmission rate of $33.04\%$, almost two times higher than that of Group 1. Group 3 was the lowest risk with an observed readmission rate of $9.85\%$ and included diagnoses for cellulitis (L03), gout (M12-19), dizziness (R42), certain fractures (S72, S22), depression (F33, F32), opioid related disorder (F11), and syncope (R55). Group 1, the largest both in number of observations and diagnoses, had an observed readmission risk of $16.87\%$, close to that of the entire cohort. 

\begin{table}[h]
  \centering 
  \caption{The number of observations and readmission rate for each diagnosis group identified by CoFA are described.} 
  \begin{tabular}{|l|c|c|}\hline
    CoFA Cluster & No. Hospitalizations & 30-day readmission rate\\ \hline 
    \ \  Group 1 & 13,003 & 16.87\% \\
    \ \ Group 2 & 1,247 & 33.04\% \\
    \ \ Group 3 & 2,843 & 9.85\% \\
    \hline 
    \ All groups & 17,093 & 16.88\% \\
    \hline 
  \end{tabular}
  \label{tbl:ccfa_stats} 
\end{table}

\subsection{Model Performance} 

\paragraph{Logistic Regression}

\begin{figure}[b]
    \centering
    \includegraphics[width=0.7\linewidth]{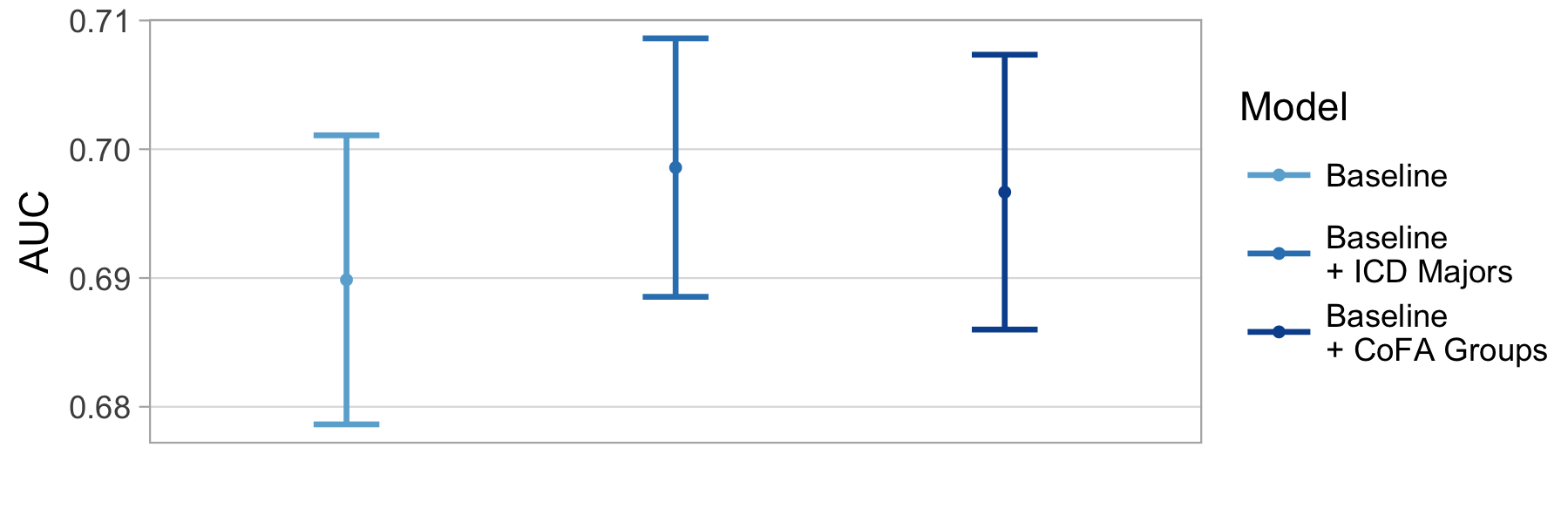}
    \caption{Three logistic regression models were fit using different categorical variables for primary diagnosis. Error bars indicate standard errors calculated by 100 iterations of repeated random sub-sampling.}
    \label{fig:simple_models}
\end{figure}

Three logistic regression models were fit using LASSO feature selection on different sets of variables: a baseline model using only non-clinical predictors, a model considering non-clinical predictors and ICD majors, and a final model considering non-clinical and CoFA categories. The baseline logistic regression model was fit considering 23 available non-diagnosis variables. Additionally considering indicator variables for the ICD major diagnosis categories improved the mean AUC on held-out test data in repeated random sub-sampling from $0.690$ to $0.699$ ($p=0.003$). Using indicator variables for the CoFA categories instead of the ICD major categories decreased the number of predictors without significantly changing the mean AUC ($\text{AUC}=0.697$, $p=0.342$). Figure \ref{fig:simple_models} compares the mean test AUCs and standard errors across 100 iterations of repeated random sub-sampling for each of the three simple logistic regression models. 

\paragraph{Custerwise Logistic Regression}

\begin{figure}[hbt]
    \centering
    \includegraphics[width=0.75\textwidth]{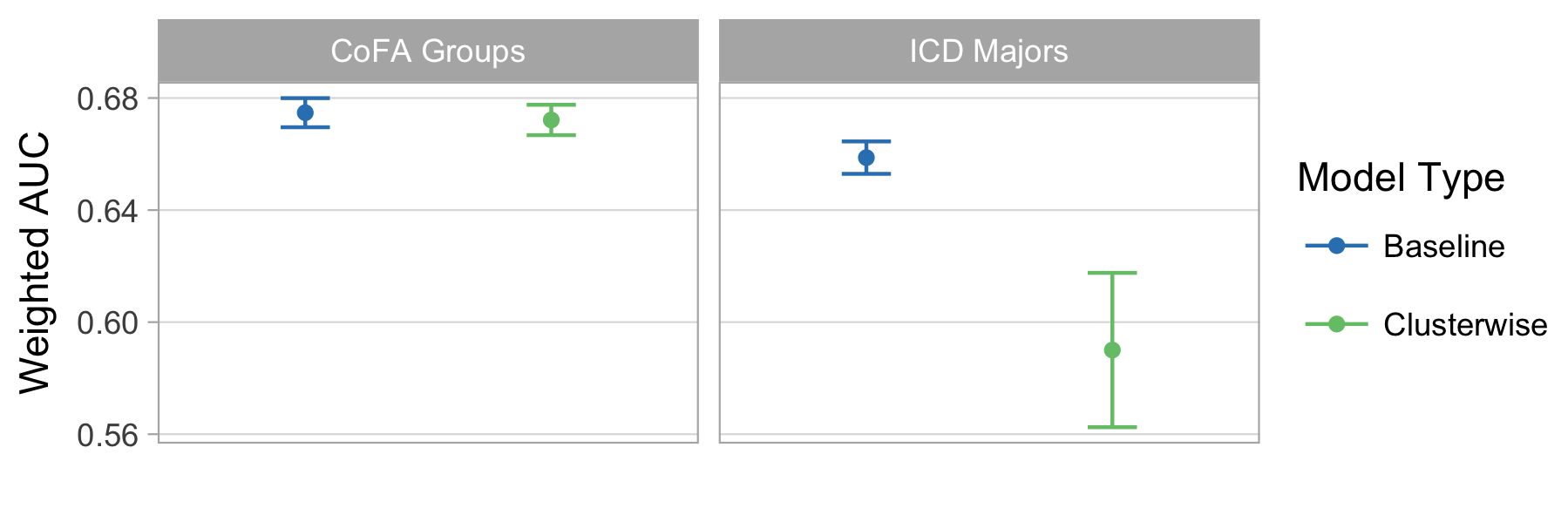}
    \caption{Models are compared using weighted AUC calculated using CoFA groups and ICD majors. For each modeling procedure the mean and standard deviation of the out-of-sample weighted AUC were calculated across 100 iterations of repeated random sub-sampling with 80\% training 20\% test splits.}
    \label{fig:cofa_icd}
\end{figure}

Two clusterwise logistic regression models were fit, stratifying by ICD majors and by CoFA groups. For each clusterwise model, the weighted AUC with respect to the categories used to stratify the clusterwise model was calculated on test data in 100 iterations of repeated random sub-sampling and then averaged to produce the mean weighted AUC. To be able to directly compare the clusterwise models to the baseline model of non-diagnosis predictors, the mean weighted AUC stratifying on the ICD majors and CoFA groups was also calculated for the baseline model. Figure \ref{fig:cofa_icd} compares the mean weighted AUC across 100 iterations of repeated random sub-sampling of each of the clusterwise models to the mean weighted AUC calculated using the same subgroups with predictions from the baseline model. 

The out-of-sample performance of the clusterwise model stratified by CoFA groups (weighted AUC $=0.672$) was not statistically significantly different ($p=0.538$) from the performance of the baseline model with non-diagnosis predictors ($\text{AUC}=0.675$). The clusterwise model using ICD majors had worse out-of-sample performance than the baseline model with mean weighted AUCs of $0.590$ and $0.659$ respectively ($p=0.029$).

\begin{figure}[htbp]
    \centering
    \includegraphics[width=0.75\textwidth]{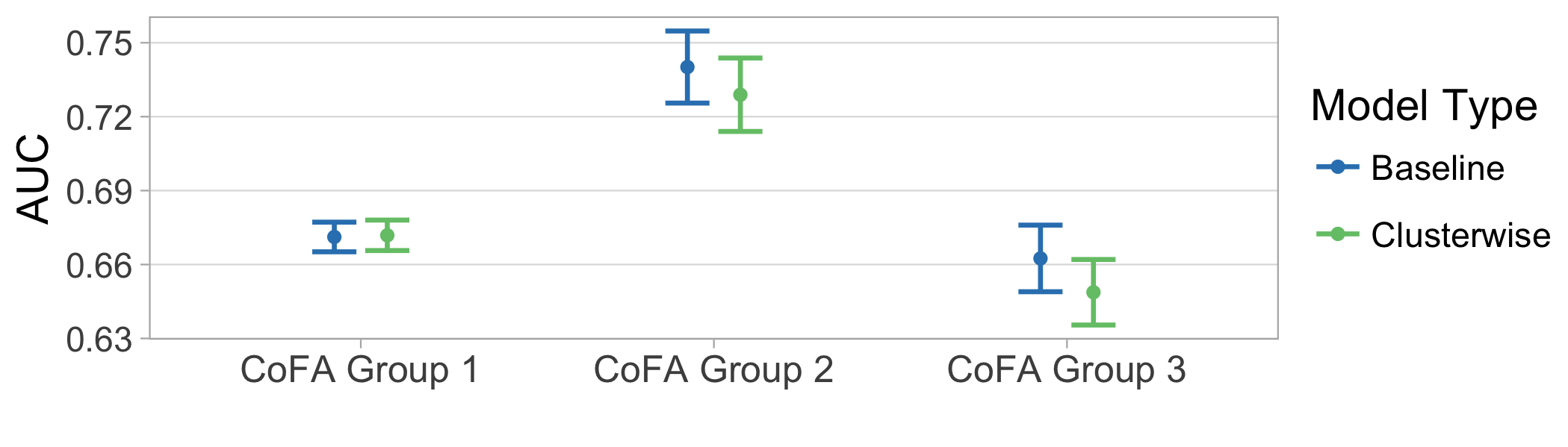}
    \caption{The out-of-sample performance of the baseline model and clusterwise model stratified by CoFA groups calculated separately for observations from each CoFA group is shown. Means and standard errors of the AUC were calculated for each group across 100 iterations of repeated random sub-sampling.}
    \label{fig:cofa_group_auc}
\end{figure}

Among the three CoFA groups, groups with higher readmission rates were easier to predict. Figure \ref{fig:cofa_group_auc} compares the AUCs calculated on predictions in each CoFA group made by the baseline model and a clusterwise model stratified by CoFA categories. Both the clusterwise and baseline model had the highest mean AUC on predictions for Group 2 ($\text{AUC}=0.672, 0.671$) among the CoFA groups. The baseline model performed similarly on both Group 1 and Group 3 with mean AUCs of $0.671$ and $0.662$, respectively. The clusterwise model did not perform as well making predictions for Group 3 ($\text{AUC}=0.649$) as when making predictions for Group 1 ($\text{AUC}=0.672$).  

\section{Discussion and Related Work} 

\subsection{Related Work}

There is considerable literature modeling variants of 30-day readmission risk, including studies comparing models fit on sub-populations to those fit to pooled data \citep{survey}. \citeauthor{yu} found that predictive models trained for individual hospitals out-performed a more general model developed on pooled data by \citeauthor{lace}, which suggests that there is variation in readmission risk factors between hospitals and patient populations \citep{yu, lace}. Fitting separate models for Diagnosis Related Groups (DRG), a system for grouping diagnoses by reimbursement rate, \citeauthor{futoma} found that for 80\% of the 260 DRG groups considered, the DRG-specific model performed better than the global model fit using the entire dataset. \citeauthor{futoma} also observed a moderate correlation between the prevalence of readmission in a DRG group and the performance of its model and a weak correlation between DRG group size and performance \citep{futoma}.

Identifying clusters of patients is advantageous for many clinical predictions problems with heterogeneous patient populations. Predicting outcomes from intensive care unit stays, \citeauthor{elbattah} used K-means to cluster observations and then fit separate random forest models that outperformed a single common model \citep{elbattah}. Algorithms introduced by \citeauthor{spath} and \citeauthor{desarbo} can perform ``clusterwise regression'' where both an optimal clustering of $k$ groups and the regression coefficients for each cluster model are estimated to minimize the overall mean squared error, but they require the prior specification of $k$ \citep{spath, desarbo}. \citeauthor{lorenzi} approached the problem of clustering clinically defined groups without pre-specifying $k$, using an agglomerate Bayesian approach called Predictive Hierarchical Clustering in which subgroups are iteratively merged to improve predictions based on Bayesian hypothesis tests \citep{lorenzi}. At the expense of interpretability, multi-task learning approaches have also been successfully used to learn models that optimize predictive power across subgroups \citep{nori, suresh, weins}. 

\subsection{Discussion}

CoFA is advantageous for quantifying similarity between levels of categorical variables because it builds upon existing CART implementations and can be used with unequally distributed categories. Our results using CoFA reveal different subgroups within the patient population defined by diagnoses that experience different levels of readmission risk: a small high risk group with alcohol-related diagnoses (Group 2), a slightly larger low risk group (Group 3) and a large group with risk similar to the overall population. Hospitalizations involving alcohol-related diagnoses (Group 2) may have higher readmission rates because of the long-term side effects of alcohol abuse. Many of the diagnoses in the lowest risk CoFA cluster (Group 3), such as dizziness (R42), fainting (R55) and fractured bones (S72, S22) were conditions that seemed unlikely to have complications or chronic effects that could cause readmission. The more chronic diagnoses clustered in group 3, such as gout (M12) and depression (F32, F33), may be associated with low 30-day readmission risk because the effects or complications occur over longer periods of time. 

The clusters identified by CoFA are useful clinical predictors of 30-day readmission. In simple logistic regression models, substituting 34 ICD majors for 3 CoFA categories simplified the model without sacrificing prediction accuracy.  Fitting separate models for each of the 34 ICD majors, had lower rank accuracy than the baseline model suggesting that ICD majors do not effectively stratify patients by readmission risk and estimating separate models for each ICD major group leads to overfitting. The clusterwise models stratified by CoFA groups performed similarly to the baseline model, suggesting that readmission risk factors may not vary enough between CoFA groups such that stratifying models by CoFA group improves predictions. For this data set, using clusterwise models stratified by ICD majors or CoFA groups did not produce more accurate predictions than using a single logistic regression model.

Evaluating the AUC on predictions for individual CoFA groups we found that the AUC of both the CoFA clusterwise model and the baseline model was higher for groups with a higher observed rate of readmission. Group 3 may have been particularly hard to model because it contained fewer positive examples of readmission and a relatively heterogeneous selection of diagnoses, compared to the other two groups. The similar performance between global models and the CoFA submodels on individual subgroups, suggest that in this dataset hospitalizations with some diagnoses are easier to model than others, whether we fit separate models for those groups or not.

There are several limitations of the data and study design used in this paper. Because we were limited to hospitalization data from Berkshire Medical Center, we could not take into account 30-day readmissions to other hospitals or patient mortality after discharge. Additionally, our sample contained multiple observations of patients who had more than one eligible index hospitalization during the study period, so not all of the observations were independent. Studies of Medicare patients have estimated 80-85\% of readmissions occur at the same hospital as the index admission \citep{nasir, bradley}. While all-hospital readmission rates provide a more accurate picture of patient outcomes, such analysis would requires extensive data to track patients between hospital systems.

There are also limitations of the CoFA method for clustering diagnoses. Our current implementation of CoFA in R uses ``dice rolling'' to take advantage of the user written split functions feature in \texttt{rpart} \citep{usersplit}. When using dice rolling to select the variables to be considered for splitting at a node, there is a non-zero probability that zero variables will be selected for consideration. We observed that for every 100 trees we fit with dice rolling, 5 would be trivial roots with no splits. Future work will include developing an R package building upon optimizations in the \texttt{randomforest} package to implement CoFA for random forests with subsetting. 


\section*{Acknowledgements} 

The authors would like to thank Gray Ellrodt and Mark Pettus of Berkshire Health Systems for providing the data, and acknowledge the support of Jeffrey Thomas and Alex Lopex of Lever Inc.


\bibliographystyle{abbrvnat}
\bibliography{mybib}


\appendix

\section*{Appendix A.}  

\begin{figure}[htb]
	\centering
	\includegraphics[width=0.9\textwidth]{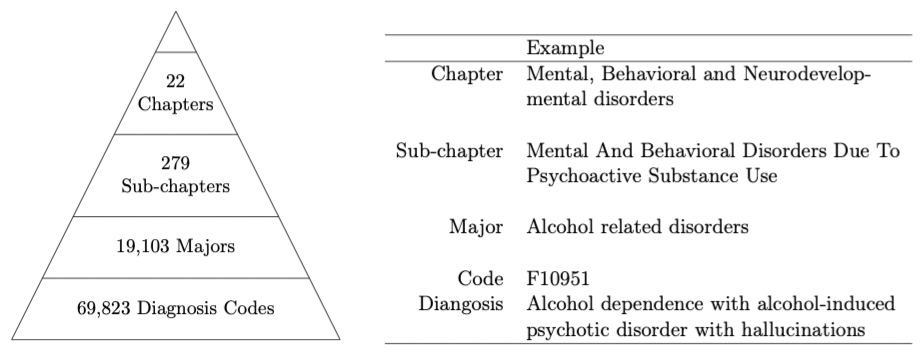}
	\caption{The pyramid shows the structure of the ICD-10-CM diagnosis coding system and the number of unique values at each level of the hierarchy. The table shows the major, sub-chapter and chapter associated with code F10951.}
	\label{fig:icd10}
\end{figure}

\begin{table}[htb]
	\caption{For each ICD major present in the sample, the CoFA group and description are shown.}
	\label{tbl:ccfa_groups}
	\centering
\begin{longtable}{|r|c|l|}
  \hline
    ICD Major & CoFA Group & Description \\ \hline
    R10 & 1 & Abdominal and pelvic pain \\ 
    Other & 1 & All other ICD majors \\ 
    N17 & 1 & Acute kidney failure \\ 
    K85 & 1 & Acute pancreatitis \\ 
    I48 & 1 & Atrial fibrillation and flutter \\ 
    F31 & 1 & Bipolar disorder \\ 
    I63 & 1 & Cerebral infarction \\ 
    K80 & 1 & Cholelithiasis \\ 
    E08 & 1 & Diabetes mellitus due to underlying condition \\ 
    K57 & 1 & Diverticular disease of intestine \\ 
    I50 & 1 & Heart failure \\ 
    J44 & 1 & Other chronic obstructive pulmonary disease \\ 
    K92 & 1 & Other diseases of digestive system \\ 
    N39 & 1 & Other disorders of urinary system \\ 
    A41 & 1 & Other sepsis \\ 
    R07 & 1 & Pain in throat and chest \\ 
    K56 & 1 & Paralytic ileus and intestinal obstruction without hernia \\ 
    J18 & 1 & Pneumonia, unspecified organism \\ 
    F43 & 1 & Reaction to severe stress, and adjustment disorders \\ 
    J96 & 1 & Respiratory failure, not elsewhere classified \\ 
    F25 & 1 & Schizoaffective disorders \\ 
    G45 & 1 & Transient cerebral ischemic attacks and related syndromes \\ 
    E11 & 1 & Type 2 diabetes mellitus \\ 
    F39 & 1 & Unspecified mood [affective] disorder \\ \hline
    F10 & 2 & Alcohol related disorders \\  \hline
   L03 & 3 & Cellulitis and acute lymphangitis \\ 
    M12 & 3 & Chronic gout \\ 
    R42 & 3 & Dizziness and giddiness \\ 
    S72 & 3 & Fracture of femur \\ 
    S22 & 3 & Fracture of rib(s), sternum and thoracic spine \\
    F33 & 3 & Major depressive disorder, recurrent \\ 
    F32 & 3 & Major depressive disorder, single episode \\ 
    F11 & 3 & Opioid related disorders \\ 
    R55 & 3 & Syncope and collapse \\ 
   \hline
\end{longtable}
\end{table}

\end{document}